\documentclass[conference]{IEEEtran}
\IEEEoverridecommandlockouts
\usepackage{lineno}
\usepackage{cite}
\usepackage{amsmath,amssymb,amsfonts}
\usepackage{amsmath}
\usepackage{amsthm}
\DeclareMathOperator*{\argmax}{argmax}
\DeclareMathOperator*{\argmin}{argmin}
\usepackage{graphicx}
\usepackage{textcomp}
\usepackage{xcolor}
\usepackage{graphicx}
\usepackage{float}
\usepackage{subfigure}
\usepackage{amsmath}
\usepackage{amsfonts,amssymb}
\usepackage{mathrsfs}
\usepackage{mathtools}
\usepackage{algorithm}
\usepackage{algorithmicx}
\usepackage{algpseudocode}
\usepackage{bm}
\usepackage{multirow}
\usepackage{array}
\usepackage{amssymb}
\usepackage{amsmath}
\usepackage{cite}
\usepackage{url}
\usepackage{xcolor}
\usepackage{cite,graphicx,amsmath,amssymb}
\usepackage{subfigure}
\usepackage{fancyhdr}
\usepackage{mdwmath}
\usepackage{mdwtab}
\usepackage{caption}
\usepackage{amsthm}
\usepackage{setspace}
\usepackage{bm}
\usepackage{algorithm}
\usepackage{algpseudocode}
\usepackage{mathtools}
\usepackage{dsfont}
\usepackage{bbm}
\newtheorem{remark}{Remark}
\newtheorem{theorem}{Theorem}

\newtheorem{lemma}{Lemma}

\newtheorem{corollary}{Corollary}

\makeatletter
\newcommand{\biggg}{\bBigg@{3}}
\newcommand{\Biggg}{\bBigg@{3.5}}
\makeatother
\def\BibTeX{{\rm B\kern-.05em{\sc i\kern-.025em b}\kern-.08em
    T\kern-.1667em\lower.7ex\hbox{E}\kern-.125emX}}
\begin{document}
\title{Aperture Selection for CAP Arrays (CAPAs)}
\author{\IEEEauthorblockN{Chongjun~Ouyang$^{\star\ddag}$, Yuanwei~Liu$^{\star}$, and Xingqi~Zhang$^{\dag}$}
$^{\star}$Queen Mary University of London, $^\ddag$University College Dublin, $^\dag$University of Alberta\\
Email: $^{\star}$\{c.ouyang, yuanwei.liu\}@qmul.ac.uk, $^{\dag}$xingqi.zhang@ualberta.ca}
\maketitle
\begin{abstract}
The concept of aperture selection is proposed for continuous aperture array (CAPA)-based communications. The achieved performance is analyzed in an uplink scenario by considering both line-of-sight (LoS) and non-line-of-sight (NLoS) scenarios. In the LoS scenario, the optimal selection strategy is demonstrated to follow the nearest neighbor criterion, and the resulting signal-to-noise ratio (SNR) is analyzed. In the NLoS scenario, the achieved outage probability along with the diversity order is revealed. Numerical results are provided to demonstrate that aperture selection effectively maintains satisfactory performance by leveraging selection diversity while simultaneously reducing the implementation complexity of CAPAs.
\end{abstract} 
\section{Introduction}
Multiple-input multiple-output (MIMO) technology has evolved into a mature technology that enhances the spectral efficiency of wireless channels. The spatial degrees of freedom (DoFs) of a MIMO system are inherently limited by the number of antennas it incorporates. To expand the available DoFs, integrating more antennas into a confined space emerges as an effective approach. It is envisaged that the ultimate evolution of existing MIMO systems will manifest as a spatially-continuous electromagnetic (EM) aperture that is characterized by an uncountable infinity of antennas separated by infinitesimal distances \cite{pizzo2020spatially}, i.e., \emph{continuous aperture array (CAPA)}. Research in this domain is still nascent; see \cite{liu2024near} and the references therein for more details.

In contrast to traditional arrays, which rely on spatially discrete (SPD) models, CAPA-based communications are rooted in continuous EM fields \cite{liu2023near}. This transition from discrete signaling in SPD arrays to continuous signaling in CAPAs introduces additional complexities in both \emph{hardware implementation and signal processing}. Furthermore, these complexities are directly proportional to the aperture size of the CAPA \cite{liu2024near}.

To achieve a better trade-off between leveraging a CAPA for enhanced communication performance and managing the associated complexity, this paper introduces the concept of \emph{aperture selection}, wherein only a portion of the continuous aperture is activated for communications. By reducing the aperture size, aperture selection enables the entire CAPA to \emph{benefit from increased spatial DoFs} while simultaneously \emph{mitigating implementation complexities}. The main contributions of this article are summarized as follows: {\romannumeral1}) We establish the analytical framework of aperture selection for CAPAs; {\romannumeral2}) Building upon the established framework, we analyze the receive signal-to-noise ratio (SNR) achieved by aperture selection in an uplink channel under both line-of-sight (LoS) and non-line-of-sight (NLoS) conditions; {\romannumeral3}) For LoS channels, we demonstrate that the optimal selection strategy adheres to the \emph{nearest neighbor criterion} and derive a closed-form expression for the corresponding SNR; {\romannumeral4})) In the case of NLoS channels, we propose a method based on discrete aperture segmentation to reduce selection complexity and characterize the resulting outage probability (OP) and diversity order; {\romannumeral5}) We demonstrate through numerical results for both scenarios that aperture selection effectively enhances system SNR by harnessing selection diversity.
\section{System Model}\label{Section: System Model}
\begin{figure}[!t]
 \centering
\setlength{\abovecaptionskip}{0pt}
\includegraphics[height=0.25\textwidth]{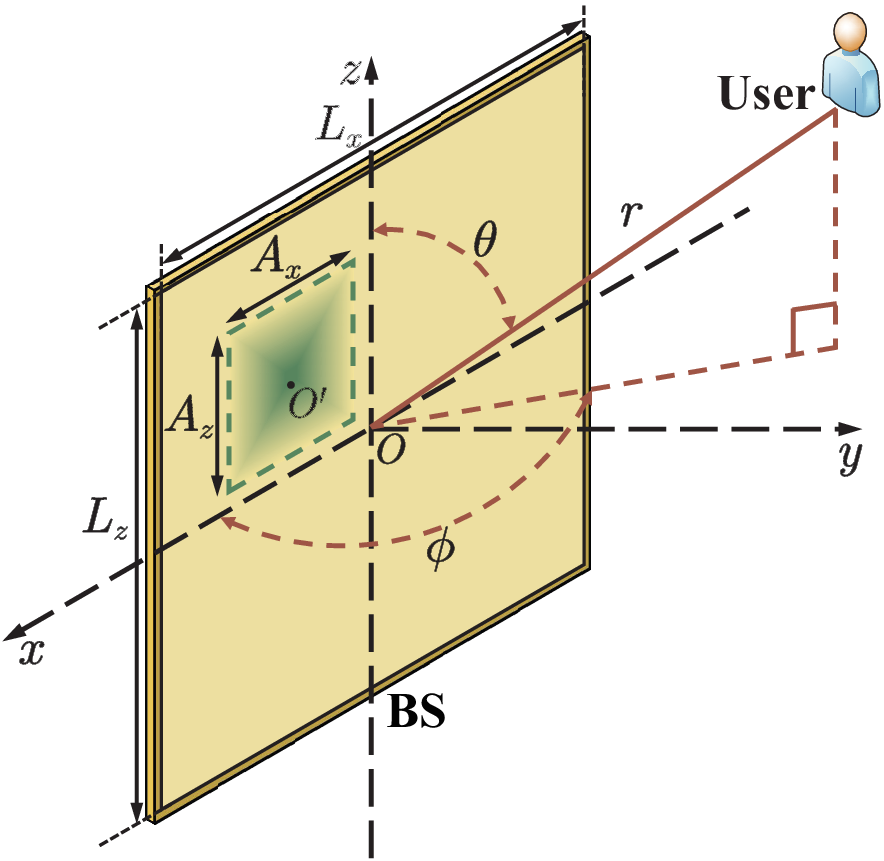}
\caption{Illustration of a CAPA.}
\label{Figure: System_Model}
\vspace{-20pt}
\end{figure}

We investigate an uplink channel where a single-antenna user transmits signals to a base station (BS) equipped with a CAPA, as shown in {\figurename} {\ref{Figure: System_Model}}. The array is placed on the $x$-$z$ plane and centered at the origin, with physical dimensions $L_x$ and $L_z$ along the $x$- and $z$-axes, respectively. Let $r$ denote the propagation distance from the center of the antenna array to the user's location, and $\phi\in[0,\pi]$ and $\theta\in[0,\pi]$ denote the associated azimuth and elevation angles, respectively. Therefore, the user's center location can be written as ${\mathbf{s}}_{\mathsf{u}}=[r\Phi,r\Psi,r\Theta]^{\mathsf{T}}$, where $\Phi\triangleq\cos{\phi}\sin{\theta}$, $\Psi\triangleq\sin\phi\sin\theta$, and $\Theta\triangleq\cos{\theta}$. The user is further assumed to employ a hypothetical isotropic antenna for signal transmission, whose transmit aperture is denoted as ${\mathcal{A}}_{\mathsf{S}}$ with $\lvert{\mathcal{A}}_{\mathsf{S}}\rvert=\frac{\lambda^2}{4\pi}$, where $\lambda$ denotes the wavelength. The BS aperture is denoted as ${\mathcal{A}}_{\mathsf{R}}=\{(x,0,z)|x\in[-\frac{L_x}{2},\frac{L_x}{2}],z\in[-\frac{L_z}{2},\frac{L_z}{2}]\}$. Throughout this paper, we assume that the aperture size of ${\mathcal{A}}_{\mathsf{R}}$ significantly exceeds that of ${\mathcal{A}}_{\mathsf{S}}$, i.e., $\lvert{\mathcal{A}}_{\mathsf{R}}\rvert\gg \lvert{\mathcal{A}}_{\mathsf{S}}\rvert$. 
\subsection{Signal Model}
Let us define ${\mathsf{J}}({\mathbf{s}})\in{\mathbbmss{C}}$ as the continuous distribution of source currents generated by the user to convey information, where ${\mathbf{s}}\in{\mathcal{A}}_{\mathsf{S}}$. The observed electric field ${\mathsf{Y}}(\mathbf{r})\in{\mathbbmss{C}}$ at point $\mathbf{r}\in{\mathcal{A}}_{\mathsf{R}}$ is the sum of the information-carrying electric field ${\mathsf{E}}(\mathbf{r})\in{\mathbbmss{C}}$ and a random noise field ${\mathsf{N}}(\mathbf{r})\in{\mathbbmss{C}}$, i.e.,
{\setlength\abovedisplayskip{2pt}
\setlength\belowdisplayskip{2pt}
\begin{align}\label{Total_electric_radiation_field}
{\mathsf{Y}}(\mathbf{r})&={\mathsf{E}}(\mathbf{r})+{\mathsf{N}}(\mathbf{r})=\int_{{\mathcal{A}}_{\mathsf{S}}}{\mathsf{h}}(\mathbf{r},{\mathbf{s}})s{\mathsf{J}}({\mathbf{s}}){\rm{d}}{\mathbf{s}}
+{\mathsf{N}}(\mathbf{r}),
\end{align}
}where $s\in{\mathbbmss{C}}$ represents the normalized symbol, ${\mathsf{N}}(\mathbf{r})$ accounts for the thermal noise, and ${\mathsf{h}}(\mathbf{r},{\mathbf{s}})\in{\mathbbmss{C}}$ denotes the spatial channel response between $\mathbf{s}$ and $\mathbf{r}$. The noise field is modeled as a zero-mean complex Gaussian random process satisfying ${\mathbbmss{E}}\{{\mathsf{N}}(\mathbf{r}_1){\mathsf{N}}^{\mathsf{H}}(\mathbf{r}_2)\}={\sigma}^2\delta(\mathbf{r}_1-\mathbf{r}_2)$, where $\delta(\cdot)$ represents the Dirac delta function.

Leveraging the fact that $\lvert{\mathcal{A}}_{\mathsf{R}}\rvert\gg \lvert{\mathcal{A}}_{\mathsf{S}}\rvert$ yields
{\setlength\abovedisplayskip{2pt}
\setlength\belowdisplayskip{2pt}
\begin{align}\label{Simplified_electric_radiation_field}
{\mathsf{Y}}(\mathbf{r})\approx s{\mathsf{h}}(\mathbf{r},{\mathbf{s}}_{\mathsf{u}}){\mathsf{J}}({\mathbf{s}}_{\mathsf{u}})\lvert{\mathcal{A}}_{\mathsf{S}}\rvert
+{\mathsf{N}}(\mathbf{r}).
\end{align}
}Subsequently, the CAPA-based BS recover $s$ from ${\mathsf{Y}}(\mathbf{r})$ using a linear combiner ${\mathsf{v}}(\mathbf{r})={\mathsf{h}}^{\mathsf{H}}(\mathbf{r},{\mathbf{s}}_{\mathsf{u}})$ along with maximum-likelihood decoding, which leads to
{\setlength\abovedisplayskip{2pt}
\setlength\belowdisplayskip{2pt}
\begin{align}\label{ML_Decoding}
\int_{{\mathcal{S}}_{\mathsf{R}}}{\mathsf{v}}(\mathbf{r}){\mathsf{Y}}(\mathbf{r}){\rm{d}}{\mathbf{r}}= s{\mathsf{a}}_{\mathsf{R}}{\mathsf{J}}({\mathbf{s}}_{\mathsf{u}})\lvert{\mathcal{A}}_{\mathsf{S}}\rvert
+\int_{{\mathcal{S}}_{\mathsf{R}}}{\mathsf{h}}^{\mathsf{H}}(\mathbf{r},{\mathbf{s}}_{\mathsf{u}}){\mathsf{N}}(\mathbf{r}){\rm{d}}{\mathbf{r}},
\end{align}
}where ${\mathsf{a}}_{\mathsf{R}}\triangleq \int_{{\mathcal{S}}_{\mathsf{R}}}\lvert{\mathsf{h}}(\mathbf{r},{\mathbf{s}}_{\mathsf{u}})\rvert^2{\rm{d}}{\mathbf{r}}>0$, and ${\mathcal{S}}_{\mathsf{R}}\subseteq{\mathcal{A}}_{\mathsf{R}}$ represent the effective surface region within the BS aperture that can capture the radiated power. Since ${\mathsf{N}}(\mathbf{r})$ is a Gaussian random field, $\int_{{\mathcal{S}}_{\mathsf{R}}}{\mathsf{h}}^{\mathsf{H}}(\mathbf{r},{\mathbf{s}}_{\mathsf{u}}){\mathsf{N}}(\mathbf{r}){\rm{d}}{\mathbf{r}}$ is complex Gaussian distributed, whose mean and variance is calculated as follows. 
\vspace{-5pt}
\begin{lemma}\label{Lemma_Noise_Distribution}
Under the assumption that ${\mathsf{N}}(\mathbf{r})$ is a zero-mean complex Gaussian process with ${\mathbbmss{E}}\{{\mathsf{N}}(\mathbf{r}_1){\mathsf{N}}^{\mathsf{H}}(\mathbf{r}_2)\}={\sigma}^2\delta(\mathbf{r}_1-\mathbf{r}_2)$, we obtain $\int_{{\mathcal{S}}_{\mathsf{R}}}{\mathsf{h}}^{\mathsf{H}}(\mathbf{r},{\mathbf{s}}_{\mathsf{u}}){\mathsf{N}}(\mathbf{r}){\rm{d}}{\mathbf{r}}\sim{\mathcal{CN}}(0,\sigma^2{\mathsf{a}}_{\mathsf{R}})$.
\end{lemma}
\vspace{-5pt}
\begin{IEEEproof}
Please refer to Appendix \ref{Proof_Lemma_Noise_Distribution} for more details.
\end{IEEEproof}
The SNR for decoding $s$ is given by
{\setlength\abovedisplayskip{2pt}
\setlength\belowdisplayskip{2pt}
\begin{align}\label{ML_Decoding_SNR}
\gamma = \frac{{\mathsf{a}}_{\mathsf{R}}^2\lvert{\mathsf{J}}({\mathbf{s}}_{\mathsf{u}})\rvert^2\lvert{\mathcal{A}}_{\mathsf{S}}\rvert^2}{\sigma^2{\mathsf{a}}_{\mathsf{R}}}
       = \frac{{\mathsf{a}}_{\mathsf{R}}\lvert{\mathsf{J}}({\mathbf{s}}_{\mathsf{u}})\rvert^2\lvert{\mathcal{A}}_{\mathsf{S}}\rvert^2}{{\sigma}^2},
\end{align}
}which is related with the effective aperture ${\mathcal{S}}_{\mathsf{R}}$. 
\subsection{Aperture Selection}
It is evident that $\gamma$ is maximized when ${\mathcal{S}}_{\mathsf{R}}={\mathcal{A}}_{\mathsf{R}}$, which, yet, leads to \emph{high complexity} in terms of \emph{hardware implementation and signal processing}, especially when ${\mathcal{A}}_{\mathsf{R}}$ has a large size. To alleviate this issue, we propose the concept of \emph{aperture selection}: selecting an aperture within ${\mathcal{A}}_{\mathsf{R}}$ to activate and using it for receiving. An example is illustrated in {\figurename} {\ref{Figure: System_Model}}, where ${\mathcal{S}}_{\mathsf{R}}$ is centered at $O'=(r_x,0,r_z)$ with dimensions $A_x\times A_z$. The corresponding selection criteria is formulated as follows:
{\setlength\abovedisplayskip{2pt}
\setlength\belowdisplayskip{2pt}
\begin{align}\label{Aperture_Selection_Criteria}
{\mathcal{S}}_{\mathsf{R}}^{\star}=\argmax_{{\mathcal{S}}_{\mathsf{R}}\subseteq{\mathcal{A}}_{\mathsf{R}}} \gamma 
= \argmax_{{\mathcal{S}}_{\mathsf{R}}\subseteq{\mathcal{A}}_{\mathsf{R}}} \int_{{\mathcal{S}}_{\mathsf{R}}}\lvert{\mathsf{h}}(\mathbf{r},{\mathbf{s}}_{\mathsf{u}})\rvert^2{\rm{d}}{\mathbf{r}},
\end{align}
}which is an NP-hard problem.
\subsection{Channel Model}
We use two typical propagation models to describe ${\mathsf{h}}(\mathbf{r},{\mathbf{s}}_{\mathsf{u}})$: {\romannumeral1}) the LoS model and {\romannumeral2}) the NLoS model.
\subsubsection{Line-of-Sight Channel}
We commence our discussion with LoS channels which can facilitate theoretical investigations into fundamental performance limits. In this case, we model ${\mathsf{h}}(\mathbf{r},{\mathbf{s}})$ as follows:
{\setlength\abovedisplayskip{2pt}
\setlength\belowdisplayskip{2pt}
\begin{align}\label{SPD_UPA_NFC_LoS_Model_User}
{\mathsf{h}}(\mathbf{r},{\mathbf{s}})={\mathsf{g}}(\mathbf{r},{\mathbf{s}})\sqrt{\frac{{\mathbf{e}}_y^{\mathsf{T}}({\mathbf{s}}-{\mathbf{r}})}{\lVert{\mathbf{r}}-{\mathbf{s}}\rVert}}\triangleq
{\mathsf{g}}_{\mathsf{LoS}}(\mathbf{r},{\mathbf{s}}),
\end{align}
}where ${\mathsf{g}}(\mathbf{r},{\mathbf{s}})\triangleq\frac{{\rm{j}}k_0\eta{\rm{e}}^{-{\rm{j}}k_0\lVert{\mathbf{r}}-{\mathbf{s}}\rVert}
}{4\pi \lVert{\mathbf{r}}-{\mathbf{s}}\rVert}$, $\eta=120\pi$ is the impedance of free space, $k_0=\frac{2\pi}{\lambda}$ is the wavenumber, and ${\mathbf{e}}_y=[0,1,0]^{\mathsf{T}}$ is the normal vector of the CAPA. Here, ${\mathsf{g}}(\mathbf{r},{\mathbf{s}})$ models the influence of free-space EM propagation, and $\frac{{\mathbf{e}}_y^{\mathsf{T}}({\mathbf{s}}-{\mathbf{r}})}{\lVert{\mathbf{r}}-{\mathbf{s}}\rVert}$ models the impact of the projected aperture of the CAPA.
\subsubsection{Non-Line-of-Sight Channel}
Next, we explore the NLoS channel model by assuming that the LoS path is blocked. In this case, we characterize ${\mathsf{h}}(\mathbf{r},{\mathbf{s}})$ as follows:
{\setlength\abovedisplayskip{2pt}
\setlength\belowdisplayskip{2pt}
\begin{align}\label{SPD_UPA_NFC_NLoS_Model_User}
{\mathsf{h}}(\mathbf{r},{\mathbf{s}})=\sum\nolimits_{n=1}^{N_{\mathsf{s}}}\alpha_{n}{\mathsf{g}}_{\mathsf{LoS}}(\mathbf{r},{\mathbf{s}}_n)
{\mathsf{g}}({\mathbf{s}}_n,{\mathbf{s}}),
\end{align}
}where $N_{\mathsf{s}}$ denotes the number of scatterers, ${\mathbf{s}}_n=[x_n,y_n,z_n]^{\mathsf{T}}$ denotes the location of the $n$th scatterer, and $\alpha_{n}\in{\mathbbmss{C}}$ denotes the associated refection coefficient. Following the Swerling-\uppercase\expandafter{\romannumeral1} model, we model $\alpha_{n}\sim{\mathcal{CN}}(0,\sigma_n^2)$ with $\sigma_n^2$ suggesting the average reflection strength.

\section{Line-of-Sight Channel}
Having established the system model, we next analyze the performance of aperture selection in a CAPA. Particularly, we commence with the LoS scenario.
\subsection{Optimal Selection Strategy}
For clarity, we assume that the activated aperture ${\mathcal{S}}_{\mathsf{R}}$ is a rectangular area. Under this assumption, we calculate the SNR in closed-form as follows.
\vspace{-5pt}
\begin{lemma}\label{Lemma_Received_SNR_Rectangular}
Let ${\mathcal{S}}_{\mathsf{R}}$ be an $A_x\times A_z$ rectangular area centered at $O'$. The SNR can be expressed follows:
{\setlength\abovedisplayskip{2pt}
\setlength\belowdisplayskip{2pt}
\begin{align}\label{Received_SNR_Rectangular}
\gamma=\frac{\overline{\gamma}}{4\pi}\sum_{x\in{\mathcal{X}}_{O'}}\sum_{z\in{\mathcal{Z}}_{O'}}\arctan\left(\frac{xz}{\Psi \sqrt{\Psi^2+x^2+z^2}}\right),
\end{align}
}where ${\mathcal{X}}_{O'}=\{\frac{A_x}{2r}\pm(\Phi-\frac{r_x}{r})\}$, ${\mathcal{Z}}_{O'}=\{\frac{A_z}{2r}\pm(\Theta-\frac{r_z}{r})\}$, and $\overline{\gamma}\triangleq\frac{\lvert{\mathsf{J}}({\mathbf{s}}_{\mathsf{u}})\rvert^2\lvert{\mathcal{A}}_{\mathsf{S}}\rvert^2k_0^2\eta^2}{4\pi{\sigma}^2}$.
\end{lemma}
\vspace{-5pt}
\begin{IEEEproof}
Please refer to Appendix \ref{Proof_Lemma_Received_SNR_Rectangular} for more details.
\end{IEEEproof}
Substituting \eqref{Received_SNR_Rectangular} into \eqref{Aperture_Selection_Criteria} gives
{\setlength\abovedisplayskip{2pt}
\setlength\belowdisplayskip{2pt}
\begin{align}\label{Aperture_Selection_Criteria_Rectangular}
{\mathcal{S}}_{\mathsf{R}}^{\star}=\argmax_{{\mathcal{S}}_{\mathsf{R}}\subseteq{\mathcal{A}}_{\mathsf{R}}} \sum_{x\in{\mathcal{X}}_{O'}}\sum_{z\in{\mathcal{Z}}_{O'}}\arctan\left(\frac{xz/\Psi}{\sqrt{\Psi^2+x^2+z^2}}\right).
\end{align}
}When ${\mathcal{S}}_{\mathsf{R}}$ is fixed to be an $A_x\times A_z$ rectangular, it is solely determined by its center location $O'$, whose feasible set is given by ${\mathcal{O}}=\{(x,0,z)|x\in[\frac{A_x-L_x}{2},\frac{L_x-A_x}{2}],z\in[\frac{A_z-L_z}{2},\frac{L_z-A_z}{2}]\}$. Therefore, the problem formulated in \eqref{Aperture_Selection_Criteria_Rectangular} can be rewritten as follows:
{\setlength\abovedisplayskip{2pt}
\setlength\belowdisplayskip{2pt}
\begin{align}\label{Aperture_Selection_Criteria_Rectangular_Trans}
\max_{(r_x,0,r_z)\in{\mathcal{O}}} \sum_{x\in{\mathcal{X}}_{O'}}\sum_{z\in{\mathcal{Z}}_{O'}}\arctan\left(\frac{xz/\Psi}{\sqrt{\Psi^2+x^2+z^2}}\right).
\end{align}
}Generally, obtaining the optimal $(r_x,0,r_z)$ requires a brute-force two-dimensional search within the feasible set ${\mathcal{O}}$, which is time-consuming. Yet, using the special mathematical structure of the SNR, Lemma \ref{Lemma_Optimal_Position_Rectangular} can be found, based on which a closed-form solution to problem \eqref{Aperture_Selection_Criteria_Rectangular_Trans} is available.
\vspace{-5pt}
\begin{lemma}\label{Lemma_Optimal_Position_Rectangular}
Given $(r,\Phi,\Theta)$, the objective function of \eqref{Aperture_Selection_Criteria_Rectangular_Trans} is monotone decreasing with either $\lvert r\Phi-r_x\rvert$ or $\lvert r\Theta-r_z\rvert$.
\end{lemma}
\vspace{-5pt}
\begin{IEEEproof}
Please refer to Appendix \ref{Proof_Lemma_Optimal_Position_Rectangular} for more details.
\end{IEEEproof}
\vspace{-5pt}
\begin{theorem}\label{Theorem_Optimal_Position_Rectangular}
Let ${\mathcal{S}}_{\mathsf{R}}$ be an $A_x\times A_z$ rectangular area centered at $O'$. The optimal $O'$ that maximizes the SNR $\gamma$ is given by
{\setlength\abovedisplayskip{2pt}
\setlength\belowdisplayskip{2pt}
\begin{align}\label{Optimal_Position_Rectangular}
{{r}}_x^{\star}&=\argmin\nolimits_{x\in[\frac{A_x-L_x}{2},\frac{L_x-A_x}{2}]}\lvert r\Phi - x\rvert,\\
{{r}}_z^{\star}&=\argmin\nolimits_{z\in[\frac{A_z-L_z}{2},\frac{L_z-A_z}{2}]}\lvert r\Theta - z\rvert,
\end{align}
}where $\argmin_{x\in[a,b]}|c-x|$ can be calculated in closed-form:
{\setlength\abovedisplayskip{2pt}
\setlength\belowdisplayskip{2pt}
\begin{align}\label{Optimal_Position_Rectangular_Closed_Form}
\argmin\nolimits_{x\in[a,b]}|c-x|=\left\{
\begin{array}{rcl}
a     &      & {c      <      a}\\
c     &      & {c\in[a,b]}\\
b     &      & {c>b}
\end{array} \right..
\end{align}
}\end{theorem}
\vspace{-5pt}
\begin{IEEEproof}
By using the result in Lemma \ref{Lemma_Optimal_Position_Rectangular}, the final result follows immediately.
\end{IEEEproof}
The function $\argmin_{x\in[a,b]}|c-x|$ returns the point in the interval $[a,b]$, which is closest to $c$. This means that $({{r}}_x^{\star},0,{{r}}_z^{\star})$ corresponds to the point in the feasible set that is closest to the projection of $\mathbf{s}_{\mathsf{u}}$ on the receive aperture, i.e., $(r\Phi,0,r\Theta)$. 
\vspace{-5pt}
\begin{remark}
The results in Theorem \ref{Theorem_Optimal_Position_Rectangular} suggest that the optimal aperture selection criterion in our considered LoS scenario is essentially the nearest neighbour criterion. This makes intuitive sense, as the nearest neighbor point minimizes both free-space path loss and projected aperture loss. 
\end{remark}
\vspace{-5pt}
\vspace{-5pt}
\begin{remark}
According to Theorem \ref{Theorem_Optimal_Position_Rectangular}, if the projection of ${\mathbf{s}}_{\mathsf{u}}$ on the receive aperture, i.e., $(r\Phi,0,r\Theta)$, is located in the feasible set $\mathcal{O}$, we have $r\Phi\in[\frac{A_x-L_x}{2},\frac{L_x-A_x}{2}]$ and $r\Theta\in[\frac{A_z-L_z}{2},\frac{L_z-A_z}{2}]$, which yields ${{r}}_x^{\star}=r\Phi$ and ${{r}}_z^{\star}=r\Theta$. This gives the performance upper bound of aperture selection, and the resulting SNR is given by
{\setlength\abovedisplayskip{2pt}
\setlength\belowdisplayskip{2pt}
\begin{align}\label{Received_SNR_Rectangular_Upper_Bound}
\gamma=\frac{\overline{\gamma}}{\pi}\arctan\left(\frac{{A_x A_z}}{2r\Psi \sqrt{4r^2\Psi^2+A_x^2+A_z^2}}\right).
\end{align}
}\end{remark}
\vspace{-5pt}
We next consider some special cases to unveil more system insights on aperture selection. 
\subsection{Further Discussions}
\subsubsection{Impact of the Selected Aperture Size}\label{Section: Impact of the Selected Aperture Size}
Let us first investigate the impact of the selected aperture size on the received SNR. For clarity, we consider the case where ${\mathcal{S}}_{\mathsf{R}}$ is a square area subject to $A_x=A_z$. We further assume that the projection of ${\mathbf{s}}_{\mathsf{u}}$ on the receive aperture, i.e., $(r\Phi,0,r\Theta)$, is located in the feasible set $\mathcal{O}$. In this case, the received SNR is given by
{\setlength\abovedisplayskip{2pt}
\setlength\belowdisplayskip{2pt}
\begin{align}\label{SNR_Optimal_Aperture_Selection}
\gamma=\frac{\overline{\gamma}}{\pi}\arctan\left({{\lvert{\mathcal{S}}_{\mathsf{R}}\rvert}}{(2r\Psi)^{-1} (4r^2\Psi^2+2\lvert{\mathcal{S}}_{\mathsf{R}}\rvert)^{-\frac{1}{2}}}\right).
\end{align}
}The SNR achieved by activating the entire receive aperture is obtained by inserting ${\mathcal{S}}_{\mathsf{R}}={\mathcal{A}}_{\mathsf{R}}$ into \eqref{SNR_Optimal_Aperture_Selection}, which yields
{\setlength\abovedisplayskip{2pt}
\setlength\belowdisplayskip{2pt}
\begin{align}
\gamma_{\mathsf{f}}=\frac{\overline{\gamma}}{\pi}\arctan\left({{\lvert{\mathcal{A}}_{\mathsf{R}}\rvert}}{(2r\Psi)^{-1} ({4r^2\Psi^2+2\lvert{\mathcal{A}}_{\mathsf{R}}\rvert})^{-\frac{1}{2}}}\right).
\end{align}
}An intriguing question to ask in this case is ``for a given $\beta\in(0,1)$, how large the active aperture size under the nearest neighbour criterion are required to achieve an SNR of $\beta \gamma_{\mathsf{f}}$?'' To this end, we calculate the derivative of $\arctan\left(\frac{x}{2r\Psi \sqrt{4r^2\Psi^2+2x}}\right)$ with respect to $x$ and note that $\frac{{\rm{d}}^2}{{\rm{d}}x}\arctan\left(\frac{x}{2r\Psi \sqrt{4r^2\Psi^2+2x}}\right)<0$ ($\forall r,\Psi$). This suggests that the achieved SNR involves a gradually decreasing incremental rate as the activated aperture size increases, which leads to the following conclusion.
\vspace{-5pt}
\begin{remark}\label{Pareto_Principle_Statement1}
A receive SNR of $\beta \gamma_{\mathsf{f}}$ is achieved by less than $\beta$ of the receive aperture ${\mathcal{A}}_{\mathsf{R}}$ being active.
\end{remark}
\vspace{-5pt}
The receive SNR in \eqref{SNR_Optimal_Aperture_Selection} can be rewritten as follows:
{\setlength\abovedisplayskip{2pt}
\setlength\belowdisplayskip{2pt}
\begin{align}\label{SNR_Optimal_Aperture_Selection_Trans1}
\gamma=\frac{\overline{\gamma}}{\pi}\arctan\left(\tau_{\mathcal{S}} {(1+2\tau_{\mathcal{S}})^{-\frac{1}{2}}}\right).
\end{align}
}where $\tau_{\mathcal{S}}\triangleq \frac{\lvert{\mathcal{S}}_{\mathsf{R}}\rvert}{4r^2\Psi^2}$. Considering the case that $\tau_{\mathcal{S}}\ll 1$, which arises when the user is located in the far-field region of the CAPA, we use the Taylor expansion of $\arctan(\cdot)$ to approximate \eqref{SNR_Optimal_Aperture_Selection_Trans1} as follows:
{\setlength\abovedisplayskip{2pt}
\setlength\belowdisplayskip{2pt}
\begin{align}\label{SNR_Optimal_Aperture_Selection_Trans2_Low}
\gamma\approx\frac{\overline{\gamma}}{\pi}\tau_{\mathcal{S}}\frac{1}{\sqrt{1+2\tau_{\mathcal{S}}}}\approx\frac{\overline{\gamma}}{\pi}\tau_{\mathcal{S}}
=\frac{\overline{\gamma}}{\pi}\frac{\lvert{\mathcal{S}}_{\mathsf{R}}\rvert}{4r^2\Psi^2},
\end{align}
}where the second approximation is due to the fact that $\frac{1}{\sqrt{1+2\tau_{\mathcal{S}}}}\approx 1$ when $\tau_{\mathcal{S}}\ll 1$. The results in \eqref{SNR_Optimal_Aperture_Selection_Trans2_Low} suggest that the receive SNR approximately scales linearly with the activated aperture size $\lvert{\mathcal{S}}_{\mathsf{R}}\rvert$ when the user is located in the far-field region, which leads to the following conclusion.
\vspace{-5pt}
\begin{remark}
The phenomenon in Remark \ref{Pareto_Principle_Statement1} is less highlighted when the user is located in the far-field region of the CAPA.
\end{remark}
\vspace{-5pt}
\subsubsection{Impact of the Aperture Shape}
We next study the impact of the aperture shape on the receive SNR. For simplicity, we focus on the case that the projection of ${\mathbf{s}}_{\mathsf{u}}$ on the receive aperture is located in the feasible region of ${\mathcal{S}}_{\mathsf{R}}$'s center location. In this case, the receive SNR achieved by activating a rectangular area is given by \eqref{SNR_Optimal_Aperture_Selection}. Following the same principle, the SNR achieved by activating a circle area with a radius of $\mathsf{R}$ can be calculated as follows:
{\setlength\abovedisplayskip{2pt}
\setlength\belowdisplayskip{2pt}
\begin{equation}\label{SNR_Optimal_Aperture_Selection_Circle}
\begin{split}
\gamma^{\prime}=\frac{\overline{\gamma}}{4\pi}\int_{0}^{2\pi}\int_{0}^{\mathsf{R}}\frac{(r\Psi)\rho {\rm{d}}\rho {\rm{d}}\theta}{(\rho^2+r^2\Psi^2)^{\frac{3}{2}}}=\frac{\overline{\gamma}}{2}\Big(1-\Big(1+\frac{4\tau_{\mathcal{S}}}{\pi}\Big)^{-\frac{1}{2}}\Big),
\end{split}
\end{equation}
}where $\tau_{\mathcal{S}}= \frac{\lvert{\mathcal{S}}_{\mathsf{R}}\rvert}{4r^2\Psi^2}=\frac{\pi{\mathsf{R}}^2}{4r^2\Psi^2}$. Let $\tau_{\mathcal{S}}\ll 1$ and obtain $\frac{4\tau_{\mathcal{S}}}{\pi}\approx0$, which, together with the Taylor expansion of $(1+x)^{-\frac{1}{2}}= 1 - \frac{1}{2}x + o(x)$, yields 
{\setlength\abovedisplayskip{2pt}
\setlength\belowdisplayskip{2pt}
\begin{equation}\label{SNR_Optimal_Aperture_Selection_Circle_Low}
\gamma^{\prime}\approx\frac{\overline{\gamma}}{2}\Big(1-\Big(1-\frac{4\tau_{\mathcal{S}}}{2\pi}\Big)\Big)=\frac{\overline{\gamma}}{\pi}\tau_{\mathcal{S}}.
\end{equation}
}We next consider the case that $\tau_{\mathcal{S}}\gg1$, which happens when the user is located in the near-field region of the CAPA. On this condition, we have
{\setlength\abovedisplayskip{2pt}
\setlength\belowdisplayskip{2pt}
\begin{equation}\label{SNR_Optimal_Aperture_Selection_Circle_Trans1}
\begin{split}
\gamma^{\prime}=\frac{\overline{\gamma}}{2}\Big(1-\Big(1+\frac{4\tau_{\mathcal{S}}}{\pi}\Big)^{-\frac{1}{2}}\Big)
\approx \frac{\overline{\gamma}}{2}\Big(1-\frac{\sqrt{\pi}}{2}\frac{1}{\sqrt{\tau_{\mathcal{S}}}}\Big).
\end{split}
\end{equation}
}By further using the expansion in \cite[Eq. (4.24.4)]{olver2010nist}, we have
{\setlength\abovedisplayskip{2pt}
\setlength\belowdisplayskip{2pt}
\begin{equation}\label{SNR_Optimal_Aperture_Selection_Trans2}
\gamma\approx\frac{\overline{\gamma}}{\pi}\Big(\frac{\pi}{2}-\frac{(1+2\tau_{\mathcal{S}})^{\frac{1}{2}}}{\tau_{\mathcal{S}}}
\Big)\approx \frac{\overline{\gamma}}{2}\Big(1-\frac{2\sqrt{2}}{\pi}\frac{1}{\sqrt{\tau_{\mathcal{S}}}}\Big)
\end{equation}
}for $\tau_{\mathcal{S}}\gg 1$. By comparing \eqref{SNR_Optimal_Aperture_Selection_Circle_Low} with \eqref{SNR_Optimal_Aperture_Selection_Trans2_Low} as well as \eqref{SNR_Optimal_Aperture_Selection_Circle_Trans1} with \eqref{SNR_Optimal_Aperture_Selection_Trans2}, we arrive at the following conclusions.
\vspace{-5pt}
\begin{remark}
The results in \eqref{SNR_Optimal_Aperture_Selection_Circle_Low} and \eqref{SNR_Optimal_Aperture_Selection_Trans2_Low} suggest that in our considered scenarios, the shape of the activated aperture has no clear impact on the receive SNR when the user is located in the far-field region. 
\end{remark}
\vspace{-5pt}
\vspace{-5pt}
\begin{remark}\label{Remark_Impact_Shape}
The results in \eqref{SNR_Optimal_Aperture_Selection_Circle_Trans1} and \eqref{SNR_Optimal_Aperture_Selection_Trans2} as well as the fact that $\frac{2\sqrt{2}}{\pi}>\frac{\sqrt{\pi}}{2}$ suggest that in the near-field region, activating a circle area achieves a receive SNR than activating a square area. However, since ${\frac{\sqrt{\pi}}{2}}/{\frac{2\sqrt{2}}{\pi}}\approx0.98$, the aforementioned SNR gap in the near-field region is quite limited.
\end{remark}
\vspace{-5pt}
\subsubsection{Discrete Aperture Segmentation}
Theorem \ref{Theorem_Optimal_Position_Rectangular} suggests that the optimal selected aperture is a function of the user's location. In practice, users may move freely, which necessitates frequent switching of the resulting optimal receive aperture ${\mathcal{S}}_{\mathsf{R}}^{\star}$. However, such frequent changes can lead to a high overhead in control signaling. One effective method to address this challenge is by discretizing the entire receive aperture into $K$ aperture segments denoted as $\{{\mathcal{S}}_{\mathsf{R}}^{(k)}\}_{k=1}^{K}\triangleq {\mathbbmss{S}}$ such that $\bigcup_{k=1}^{K}{\mathcal{S}}_{\mathsf{R}}^{(k)}={\mathcal{A}}_{\mathsf{R}}$, and selecting the aperture from this candidate set. Under this circumstance, the optimal aperture index is determined as follows:
{\setlength\abovedisplayskip{2pt}
\setlength\belowdisplayskip{2pt}
\begin{equation}
k^{\star}=\argmax\limits_{k=1,\ldots,K}\int_{{\mathcal{S}}_{\mathsf{R}}^{(k)}}\lvert{\mathsf{h}}(\mathbf{r},{\mathbf{s}}_{\mathsf{u}})\rvert^2{\rm{d}}{\mathbf{r}}.
\end{equation}
}By definition, we obtain
{\setlength\abovedisplayskip{2pt}
\setlength\belowdisplayskip{2pt}
\begin{equation}
\int_{{\mathcal{S}}_{\mathsf{R}}^{(k^{\star})}}\lvert{\mathsf{h}}(\mathbf{r},{\mathbf{s}}_{\mathsf{u}})\rvert^2{\rm{d}}{\mathbf{r}}
\leq \int_{{\mathcal{S}}_{\mathsf{R}}^{\star}}\lvert{\mathsf{h}}(\mathbf{r},{\mathbf{s}}_{\mathsf{u}})\rvert^2{\rm{d}}{\mathbf{r}},
\end{equation}
}which indicates the sub-optimality of the aperture segmentation. A simple aperture segmentation strategy involves partitioning the receive aperture into four disjoint parts, each corresponding to a quadrant. Specifically, we have 
{\setlength\abovedisplayskip{2pt}
\setlength\belowdisplayskip{2pt}
\begin{equation}
{\mathbbmss{S}}=\{\{(\pm x,0,\pm z)|x \in[0,{L_x}/{2}],z\in[0,{L_z}/{2}]\}\}\triangleq {\mathbbmss{S}}_{\mathsf{q}},\nonumber
\end{equation}
}which yields $O'\in\{(\pm {L_x}/{4},0,\pm {L_z}/{4})\}$. By using the results in Theorem \ref{Theorem_Optimal_Position_Rectangular}, the following conclusion is drawn.
\vspace{-5pt}
\begin{remark}
When ${\mathbbmss{S}}={\mathbbmss{S}}_{\mathsf{q}}$, the optimal candidate aperture that can maximize the receive SNR is the quadrant where the projection of $\mathbf{s}_{\mathsf{u}}$ on ${\mathcal{A}}_{\mathsf{R}}$ is located in.
\end{remark}
\vspace{-5pt}
\subsubsection{Linear Array}
We consider the case when $\frac{L_z}{r}\ll 1$, where the variation of the channel response along the $z$-axis is negligible. This makes the entire \emph{planar array} degenerate into a \emph{linear array} along the $x$-axis. The receive SNR achieved by selecting the interval $[r_x-\frac{A_x}{2},r_x+\frac{A_x}{2}]$ is given by
{\setlength\abovedisplayskip{2pt}
\setlength\belowdisplayskip{2pt}
\begin{align}
\zeta&=\overline{\gamma}\frac{A_z}{4\pi}\int_{r_x-\frac{A_x}{2}}^{r_x+\frac{A_x}{2}}
\frac{r{\Psi}{\rm{d}}x}{((x-r\Phi)^2+r^2{\Psi}^2+r^2\Theta^2)^{\frac{3}{2}}}\\
&=\left.\frac{\overline{\gamma}\frac{A_z}{4\pi}\frac{x-r_{x,\Phi}}{r^2{\Psi}^2+r^2\Theta^2}r{\Psi}}{\sqrt{(x-r_{x,\Phi})^2+r^2{\Psi}^2+r^2\Theta^2}}\right\rvert_{-A_x/2}^{A_x/2}
\end{align}
}with $r_{x,\Phi}= r\Phi-r_x$. Following the steps in obtaining Theorem \ref{Theorem_Optimal_Position_Rectangular}, the following corollary can be found. 
\vspace{-5pt}
\begin{corollary}\label{Corollary_Optimal_Position_Linear}
Given the above linear array, the optimal $r_x$ that maximizes the SNR $\gamma$ is given by
{\setlength\abovedisplayskip{2pt}
\setlength\belowdisplayskip{2pt}
\begin{align}
{{r}}_x^{\star}&=\argmin\nolimits_{x\in[\frac{A_x-L_x}{2},\frac{L_x-A_x}{2}]}\lvert r\Phi - x\rvert,
\end{align}
}where $\argmin_{x\in[a,b]}|c-x|$ is calculated by \eqref{Optimal_Position_Rectangular_Closed_Form}.
\end{corollary}
\vspace{-5pt}
The function $\argmin_{x\in[a,b]}|c-x|$ returns the point in the interval $[a,b]$, which is closest to $c$. Thus, ${{r}}_x^{\star}$ corresponds to the point in the feasible set $[\frac{A_x-L_x}{2},\frac{L_x-A_x}{2}]$, which is closest to the projection of $\mathbf{s}_{\mathsf{u}}$ on the linear array, i.e., $r\Phi$. 
\vspace{-5pt}
\begin{remark}
The results in Corollary \ref{Corollary_Optimal_Position_Linear} suggest that the optimal interval selection criterion in our considered scenario is the nearest neighbour criterion, which is akin to Theorem \ref{Theorem_Optimal_Position_Rectangular}. 
\end{remark}
\vspace{-5pt}
\section{Non-Line-of-Sight Channel}
\subsection{Optimal Aperture Selection}
Having analyzed the performance of aperture selection for the LoS channel, we now shift our attention to NLoS channels. For the sake of brevity, we concentrate our attention on the case where the activated aperture ${\mathcal{S}}_{\mathsf{R}}$ is an $A_x\times A_z$ rectangular area centered at $(r_x,0,r_z)$. Given ${\mathcal{S}}_{\mathsf{R}}$, the receive SNR under the NLoS channel model shown in \eqref{SPD_UPA_NFC_NLoS_Model_User} is calculated as follows:
{\setlength\abovedisplayskip{2pt}
\setlength\belowdisplayskip{2pt}
\begin{equation}\label{Received_SNR_Rectangular_NLoS}
\begin{split}
&\gamma_{\mathsf{NLoS}}={\overline{\gamma}}\sum\nolimits_{n=1}^{N_{\mathsf{s}}}\lvert\alpha_{n}\rvert^2
\lvert{\mathsf{g}}({\mathbf{s}}_n,{\mathbf{s}}_{\mathsf{u}})\rvert^2{\mathsf{a}}_{\mathsf{LoS}}^{(n)}+{\overline{\gamma}}\sum\nolimits_{n=1}^{N_{\mathsf{s}}}\\
&\times\sum\nolimits_{n'\ne n}\alpha_{n}\alpha_{n'}^{\mathsf{H}}
{\mathsf{g}}({\mathbf{s}}_n,{\mathbf{s}}_{\mathsf{u}}){\mathsf{g}}^{\mathsf{H}}({\mathbf{s}}_{n'},{\mathbf{s}}_{\mathsf{u}})\sqrt{y_ny_{n'}}
{{\rho}}_{n,n'},
\end{split}
\end{equation}
}where ${\mathsf{a}}_{\mathsf{LoS}}^{(n)}\triangleq\frac{1}{4\pi}\int_{{\mathcal{S}}_{\mathsf{R}}}
\frac{y_n{\rm{d}}x{\rm{d}}z}{((x-x_n)^2+y_n^2+(z-z_n)^2)^{{3}/{2}}}$ and ${{\rho}}_{n,n'}\triangleq\frac{1}{4\pi}\int_{{\mathcal{S}}_{\mathsf{R}}}
\frac{{\rm{e}}^{{\rm{j}}k_0(({(x-x_{n'})^2+y_{n'}^2+(z-z_{n'})^2})^{\frac{1}{2}}-({(x-x_n)^2+y_n^2+(z-z_n)^2})^{\frac{1}{2}})}{\rm{d}}x{\rm{d}}z}
{((x-x_n)^2+y_n^2+(z-z_n)^2)^{{3}/{4}}((x-x_{n'})^2+y_{n'}^2+(z-z_{n'})^2)^{{3}/{4}}}$. Note that ${\mathsf{a}}_{\mathsf{LoS}}^{(n)}$ can be calculated in closed-form as \eqref{Received_SNR_Rectangular}, whereas ${{\rho}}_{n,n'}$ can be approximated using the Gauss–Legendre quadrature rule \cite{olver2010nist} twice, which yields
{\setlength\abovedisplayskip{2pt}
\setlength\belowdisplayskip{2pt}
\begin{align}
&{{\rho}}_{n,n'}\approx\sum_{i=1}^{T}\sum_{j=1}^{T}
\frac{{\rm{e}}^{{\rm{j}}k_0((\frac{A_x\xi_j}{2}+r_{x,n'})^2+y_{n'}^2+(\frac{A_z\xi_i}{2}+r_{z,n'})^2)^{\frac{1}{2}}}}
{((\frac{A_x\xi_j}{2}+r_{x,n'})^2+y_{n'}^2+(\frac{A_z\xi_i}{2}+r_{z,n'})^2)^{\frac{3}{4}}}
\nonumber\\
&~~\times\frac{\frac{w_iw_j}{4\pi}\frac{A_xA_z}{4}{\rm{e}}^{-{\rm{j}}k_0((\frac{A_x\xi_j}{2}+r_{x,n})^2+y_n^2+(\frac{A_z\xi_i}{2}+r_{z,n})^2)^{\frac{1}{2}}}}
{((\frac{A_x\xi_j}{2}+r_{x,n})^2+y_n^2+(\frac{A_z\xi_i}{2}+r_{z,n})^2)^{\frac{3}{4}}},
\end{align}
}where $r_{x,n}\triangleq r_x-x_n$, $r_{z,n}\triangleq r_z-z_n$, where $\{w_i\}$ and $\{\xi_i\}$ denote the weight and abscissa factors of Gauss–Legendre integration, and $T$ is a complexity-vs-accuracy tradeoff parameter. Under the NLoS scenario, the aperture selection criteria is formulated as follows:
{\setlength\abovedisplayskip{2pt}
\setlength\belowdisplayskip{2pt}
\begin{align}\label{Aperture_Selection_Criteria_NLoS}
{\mathcal{S}}_{\mathsf{R}}^{\star}=\argmax\nolimits_{{\mathcal{S}}_{\mathsf{R}}\subseteq{\mathcal{A}}_{\mathsf{R}}}\gamma_{\mathsf{NLoS}}.
\end{align}
}Upon comparing \eqref{Received_SNR_Rectangular} with \eqref{Received_SNR_Rectangular_NLoS}, we observe that the receive SNR in the NLoS scenario is more intractable than that in the LoS scenario due to the existence of multiple scatterers. 
\vspace{-5pt}
\begin{remark}
The intractability of \eqref{Received_SNR_Rectangular_NLoS} makes the nearest neighbour criterion not optimal for the NLoS channels.
\end{remark}
\vspace{-5pt} 
\subsection{Discrete Aperture Segmentation}
Finding the optimal solution to problem \eqref{Aperture_Selection_Criteria_NLoS} asks for a brute-force two-dimensional search within the receive aperture ${\mathcal{A}}_{\mathsf{R}}$, which is computationally intensive. As a compromise, we propose to portion the entire aperture into several segments $\{{\mathcal{S}}_{\mathsf{R}}^{(k)}\}_{k=1}^{K}= {\mathbbmss{S}}$, and selecting the aperture from this candidate set, i.e., $k^{\star}=\argmax_{k=1,\ldots,K}\gamma_k$, where $\gamma_k\triangleq\int_{{\mathcal{S}}_{\mathsf{R}}^{(k)}}\lvert\sum\nolimits_{n=1}^{N_{\mathsf{s}}}\alpha_{n}{\mathsf{g}}_{\mathsf{LoS}}(\mathbf{r},{\mathbf{s}}_n)
{\mathsf{g}}({\mathbf{s}}_n,{\mathbf{s}})\rvert^2{\rm{d}}{\mathbf{r}}$. Since $\alpha_{n}\sim{\mathcal{CN}}(0,\sigma_n^2)$ for $n=1,\ldots,N_{\mathsf{s}}$, we have $\gamma_k\overset{d}{=}\lvert h_k\rvert$, where $\overset{d}{=}$ denotes the equality in distribution and $h_k$ is complex Gaussian distributed. By defining ${\mathbf{h}}\triangleq[h_1,\ldots,h_K]^{\mathsf{T}}$, we have ${\mathbf{h}}\sim{\mathcal{CN}}({\mathbf{0}},{\mathbf{R}})$ where ${\mathbf{R}}={\mathbbmss{E}}\{{\mathbf{h}}{\mathbf{h}}^{\mathsf{H}}\}$ represents the correlation matrix. To describe $\mathbf{R}$, we introduce the following lemma.
\vspace{-5pt}
\begin{lemma}\label{Lemma_Correlation_Operator}
Let ${\mathbf{r}}_1,{\mathbf{r}}_2\in{\mathcal{A}}_{\mathsf{R}}$. The correlation function ${\mathscr{R}}({\mathbf{r}}_1,{\mathbf{r}}_2)\triangleq{\mathbbmss{E}}\{\int_{{\mathcal{A}}_{\mathsf{S}}}{\mathsf{h}}(\mathbf{r}_1,{\mathbf{s}})
{\mathsf{h}}^{\mathsf{H}}(\mathbf{r}_2,{\mathbf{s}}){\rm{d}}{\mathbf{s}}\}$ for the NLoS spatial response \eqref{SPD_UPA_NFC_NLoS_Model_User} is given by
{\setlength\abovedisplayskip{2pt}
\setlength\belowdisplayskip{2pt}
\begin{equation}
\begin{split}
&{\mathscr{R}}({\mathbf{r}}_1,{\mathbf{r}}_2)={\overline{\gamma}}\sum\nolimits_{n=1}^{N_{\mathsf{s}}}
\sigma_n^2\lvert{\mathsf{g}}({\mathbf{s}}_n,{\mathbf{s}}_{\mathsf{u}})\rvert^2
\frac{{\rm{e}}^{-{\rm{j}}k_0\lVert{\mathbf{r}}_1-{\mathbf{s}}_n\rVert}}{\sqrt{4\pi} \lVert{\mathbf{r}}_1-{\mathbf{s}}_n\rVert}\\
&\times\sqrt{\frac{{\mathbf{e}}_y^{\mathsf{T}}({\mathbf{s}}_n-{\mathbf{r}}_1)}{\lVert{\mathbf{r}}_1-{\mathbf{s}}_n\rVert}}
\frac{{\rm{e}}^{{\rm{j}}k_0\lVert{\mathbf{r}}_2-{\mathbf{s}}_n\rVert}}{\sqrt{4\pi} \lVert{\mathbf{r}}_2-{\mathbf{s}}_n\rVert}\sqrt{\frac{{\mathbf{e}}_y^{\mathsf{T}}({\mathbf{s}}_n-{\mathbf{r}}_2)}{\lVert{\mathbf{r}}_2-{\mathbf{s}}_n\rVert}}.
\end{split}
\end{equation}
}\end{lemma}
\vspace{-5pt}
\begin{IEEEproof}
This lemma is proved by substituting \eqref{SPD_UPA_NFC_NLoS_Model_User} into the expression of ${\mathscr{R}}({\mathbf{r}}_1,{\mathbf{r}}_2)$ and using the fact that ${\mathbbmss{E}}\{\lvert\alpha_{n}\rvert^2\}=\sigma_n^2$ for $n=1,\ldots,N_{\mathsf{s}}$ and ${\mathbbmss{E}}\{\alpha_{n}^{\mathsf{H}}\alpha_{n'}\}=0$ for $n\ne n'$.
\end{IEEEproof}
After obtaining ${\mathscr{R}}({\mathbf{r}}_1,{\mathbf{r}}_2)$, the correlation matrix $\mathbf{R}$ can be directly constructed. Particularly, the $(k,k')$th element of $\mathbf{R}$ can be calculated as follows:
{\setlength\abovedisplayskip{2pt}
\setlength\belowdisplayskip{2pt}
\begin{equation}
\begin{split}\label{Correlation_Matrix_Element_Basic}
[\mathbf{R}]_{k,k'}=\int_{{\mathcal{S}}_{\mathsf{R}}^{(k)}}\int_{{\mathcal{S}}_{\mathsf{R}}^{(k')}}{\mathscr{R}}({\mathbf{r}}_1,{\mathbf{r}}_2){\rm{d}}{\mathbf{r}}_1{\rm{d}}{\mathbf{r}}_2.
\end{split}
\end{equation}
}We comment that the double integral in \eqref{Correlation_Matrix_Element_Basic} can be approximated using the Gauss–Legendre quadrature rule, which is akin to \eqref{Received_SNR_Rectangular_NLoS}. In summary, the following lemma is found.
\vspace{-5pt}
\begin{lemma}
Under the considered model, it has $\gamma_{k^{\star}}\overset{d}{=}\max_{k=1,\ldots,K}\lvert h_k\rvert^2$ with $[h_1,\ldots,h_K]^{\mathsf{T}}={\mathbf{h}}\sim{\mathcal{CN}}({\mathbf{0}},{\mathbf{R}})$.
\end{lemma}
\vspace{-5pt}
Since ${\mathbf{h}}\sim{\mathcal{CN}}({\mathbf{0}},{\mathbf{R}})$, we have ${\mathbf{h}}\overset{d}{=}{\mathbf{R}}^{\frac{1}{2}}\tilde{\mathbf{h}}$ with $\tilde{\mathbf{h}}\sim{\mathcal{CN}}({\mathbf{0}},{\mathbf{I}}_K)$. We next characterize the OP of $\gamma_{k^{\star}}$, which is defined as $\Pr(\gamma_{k^{\star}}<\gamma_{\mathsf{th}})$ with $\gamma_{\mathsf{th}}$ representing the target SNR. Let ${\mathbf{U}}^{\mathsf{H}}{\bm\Lambda}{\mathbf{U}}$ denote the eigendecomposition of ${\mathbf{R}}$, where ${\mathbf{U}}\in{\mathbbmss{C}}^{{\mathsf{r}}\times {K}}$ is a semi-unitary matrix with ${\mathbf{U}}{\mathbf{U}}^{\mathsf{H}}={\mathbf{I}}_{\mathsf{r}}$, and ${\bm\Lambda}=\mathsf{diag}([\lambda_1,\ldots,\lambda_{\mathsf{r}}]^{\mathsf{T}})\in{\mathbbmss{C}}^{{\mathsf{r}}\times {\mathsf{r}}}$ with $\{\lambda_{i}\}_{i=1}^{\mathsf{r}}$ being the positive eigenvalues of $\mathbf{R}$, and $\mathsf{r}$ is the rank of $\mathbf{R}$. The statistical distribution of $\mathbf{h}$ is thus given by \cite[pp. 527--528]{rao1973linear}
{\setlength\abovedisplayskip{2pt}
\setlength\belowdisplayskip{2pt}
\begin{equation}
f_{\mathbf{h}}(\mathbf{x})={\rm{e}}^{-{\mathbf{x}}^{\mathsf{H}}{\mathbf{R}}^{+}{\mathbf{x}}}/({{\pi^{\mathsf{r}}{{\det}^{*}}({\mathbf{R}})}}),
\end{equation}
}where ${\mathbf{R}}^{+}={\mathbf{U}}^{\mathsf{H}}{\bm\Lambda}^{-1}{\mathbf{U}}\in{\mathbbmss{C}}^{K\times K}$ is the generalized inverse of $\mathbf{R}$, and $\det^{*}({\mathbf{R}})=\prod_{i=1}^{\mathsf{r}}\lambda_i$ is the pseudo-determinant of $\mathbf{R}$. Using the moment generating function-based strategy, an exact expression for $\Pr(\gamma_{k^{\star}}<\gamma_{\mathsf{th}})$ is available, which is yet quite articulated and does not provide important insights. As a compromise, we focus on the high-SNR asymptotic behavior of $\Pr(\gamma_{k^{\star}}<\gamma_{\mathsf{th}})$ by letting $\overline{\gamma}\rightarrow\infty$. 
\vspace{-5pt}
\begin{theorem}\label{Theorem_Correlation_Outage_Probability}
Let $\overline{\gamma}\rightarrow\infty$, and the OP satisfies 
{\setlength\abovedisplayskip{2pt}
\setlength\belowdisplayskip{2pt}
\begin{equation}
\Pr(\gamma_{k^{\star}}<\gamma_{\mathsf{th}})={\mathcal{O}}(\overline{\gamma}^{-{\mathsf{r}}}\gamma_{\mathsf{th}}^{\mathsf{r}}/{{{\det}^{*}}({\mathbf{R}})}).
\end{equation}
}\end{theorem}
\vspace{-5pt}
\begin{IEEEproof}
Please refer to \cite{liu2009large,liu2010asymptotic} for more details.
\end{IEEEproof}
\vspace{-5pt}
\begin{remark}
Theorem \ref{Theorem_Correlation_Outage_Probability} suggests that under the NLoS condition, the diversity order achieved by discrete aperture selection is given by $\mathsf{r}$, i.e., the rank of the correlation matrix $\mathbf{R}$.
\end{remark}
\vspace{-5pt}
According to the properties of the rank of a matrix, we have ${\mathsf{r}}\leq \min\{K,N_{\mathsf{s}}\}$. This means that the diversity order is mainly limited by the number of scatterers in a poorly scattered environment where $N_{\mathsf{s}}$ is a finite value. By contrast, for a richly scattered environment where $N_{\mathsf{s}}=\infty$, the diversity order will be mainly limited by $K$, i.e., the size of the candidate set $\mathbbmss{S}$.
\section{Numerical Results}
In this section, we employ numerical results to validate the effectiveness of aperture selection. Unless explicitly stated otherwise, we set the following parameters: $\overline{\gamma}=40$ dB, $\theta=\frac{\pi}{3}$, $\phi=\frac{\pi}{6}$, $r=10$ m, $\lambda = 0.0107$ m, $A_x=A_z$, and $L_x=L_z$.
\begin{figure}[!t]
    \centering
    \subfigbottomskip=0pt
	\subfigcapskip=-5pt
\setlength{\abovecaptionskip}{0pt}
    \subfigure[$A_x=L_x/8$.]
    {
        \includegraphics[height=0.17\textwidth]{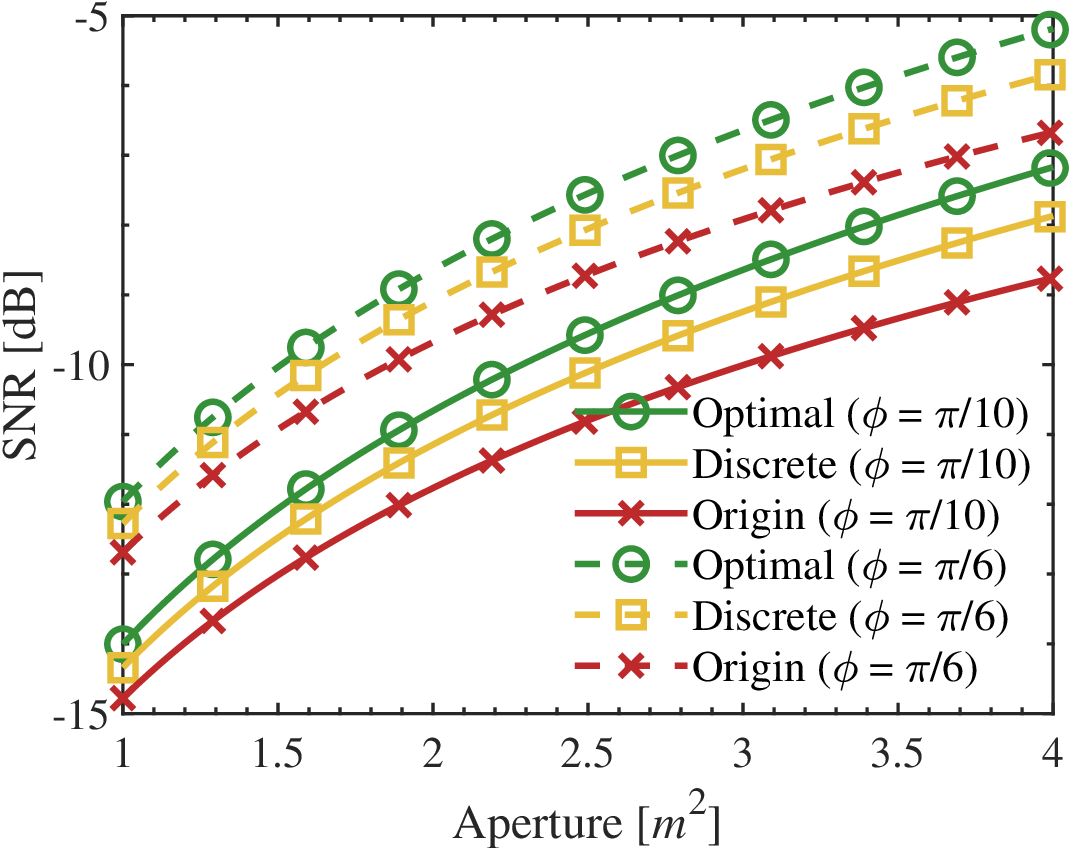}
	   \label{fig2a}	
    }
   \subfigure[$L_x=2$ m.]
    {
        \includegraphics[height=0.17\textwidth]{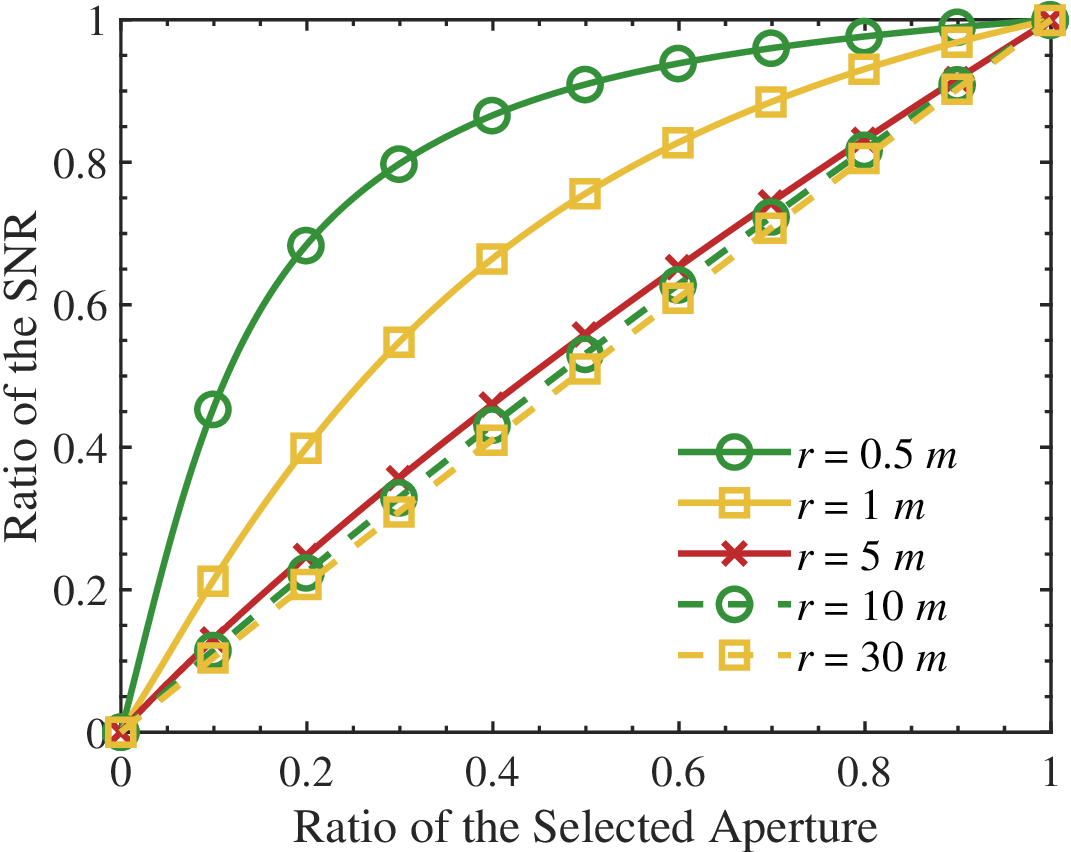}
	   \label{fig2b}	
    }
\caption{Aperture selection for LoS channels.}
    \label{Figure2}
    \vspace{-10pt}
\end{figure}

{\figurename} {\ref{fig2a}} illustrates the receive SNR for LoS channels plotted against the aperture size $\lvert {\mathcal{A}}_{\mathsf{R}}\rvert$ when the selected aperture is an $A_x\times A_x$ square. For comparison, we present the results achieved by the optimal selection strategy (the nearest neighbor criterion), discrete aperture segmentation, and no selection (${\mathcal{S}}_{\mathsf{R}}$ fixed to be centered at the origin). In the discrete aperture segmentation, we consider a finite candidate containing five square areas centered at ${\mathbbmss{o}}\triangleq\{[0,0,0]^{\mathsf{T}},[\pm\frac{L_x}{2},0,\pm\frac{L_z}{2}]^{\mathsf{T}}\}$. It is observed that both proposed aperture selection methods effectively enhance the receive SNR compared to the scheme without aperture selection. This improvement stems from the utilization of selection diversity. To further explore the effectiveness of aperture selection, {\figurename} {\ref{fig2b}} investigates the impact of the selected aperture size $\lvert {\mathcal{S}}_{\mathsf{R}}\rvert$, where the selection is based on the nearest neighbor criterion and the entire aperture size is fixed to $\lvert {\mathcal{A}}_{\mathsf{R}}\rvert=4~{\text{m}}^2$. {\figurename} {\ref{fig2b}} depicts $\frac{\gamma}{\gamma_{\mathsf{f}}}$ (the ratio of the SNR) as a function of $\frac{\lvert {\mathcal{S}}_{\mathsf{R}}\rvert}{\lvert {\mathcal{A}}_{\mathsf{R}}\rvert}$ (the ratio of the aperture size) for selected values of $r$, where ${\gamma_{\mathsf{f}}}$ represents the SNR achieved by setting $\frac{\lvert {\mathcal{S}}_{\mathsf{R}}\rvert}{\lvert {\mathcal{A}}_{\mathsf{R}}\rvert}=1$. As shown, for small values of $r$, i.e., when the user is located in near-field region, 60\% of ${\gamma_{\mathsf{f}}}$ is achieved by less than 40\% of the receive aperture being active. However, as $r$ increases, $\frac{\gamma}{\gamma_{\mathsf{f}}}$ approximately increases linearly with $\frac{\lvert {\mathcal{S}}_{\mathsf{R}}\rvert}{\lvert {\mathcal{A}}_{\mathsf{R}}\rvert}$. These results align with our discussion in Section \ref{Section: Impact of the Selected Aperture Size}, which suggests that aperture selection is more efficient in the near-field region for LoS channels.

\begin{figure}[!t]
    \centering
    \subfigbottomskip=0pt
	\subfigcapskip=-5pt
\setlength{\abovecaptionskip}{0pt}
    \subfigure[Achieved SNRs for various shapes.]
    {
        \includegraphics[height=0.17\textwidth]{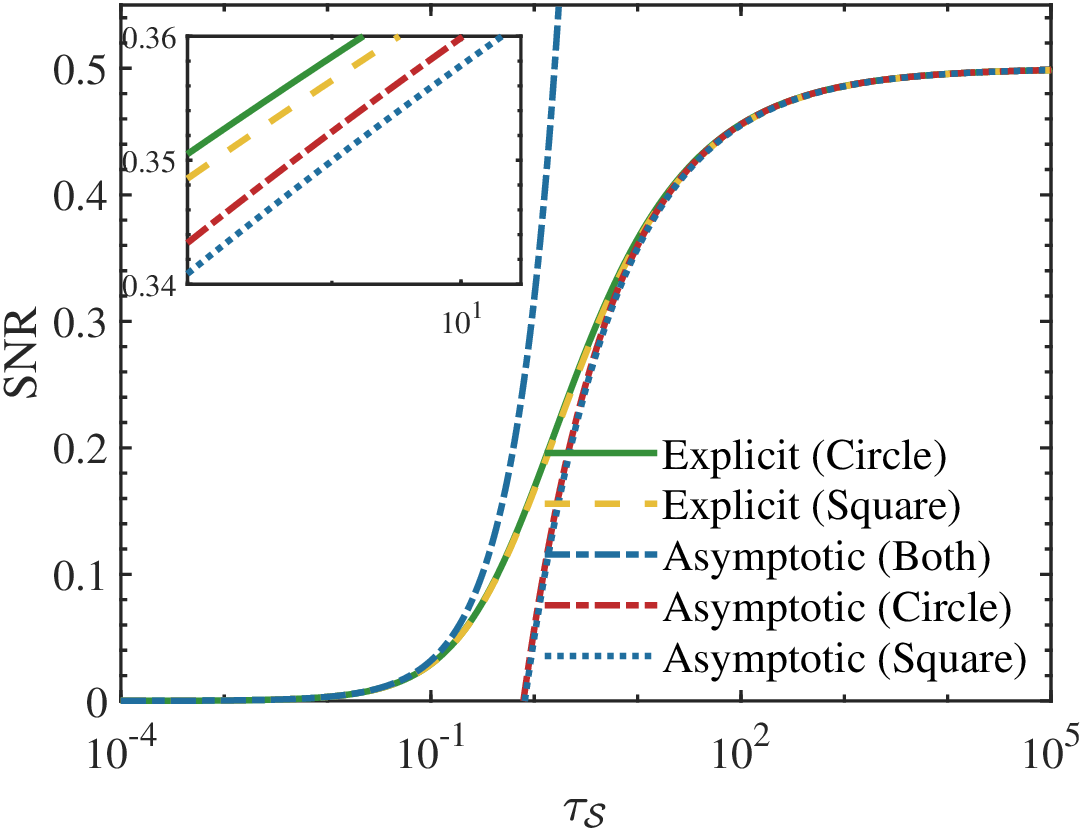}
	   \label{fig3a}	
    }
   \subfigure[SNR ratio.]
    {
        \includegraphics[height=0.17\textwidth]{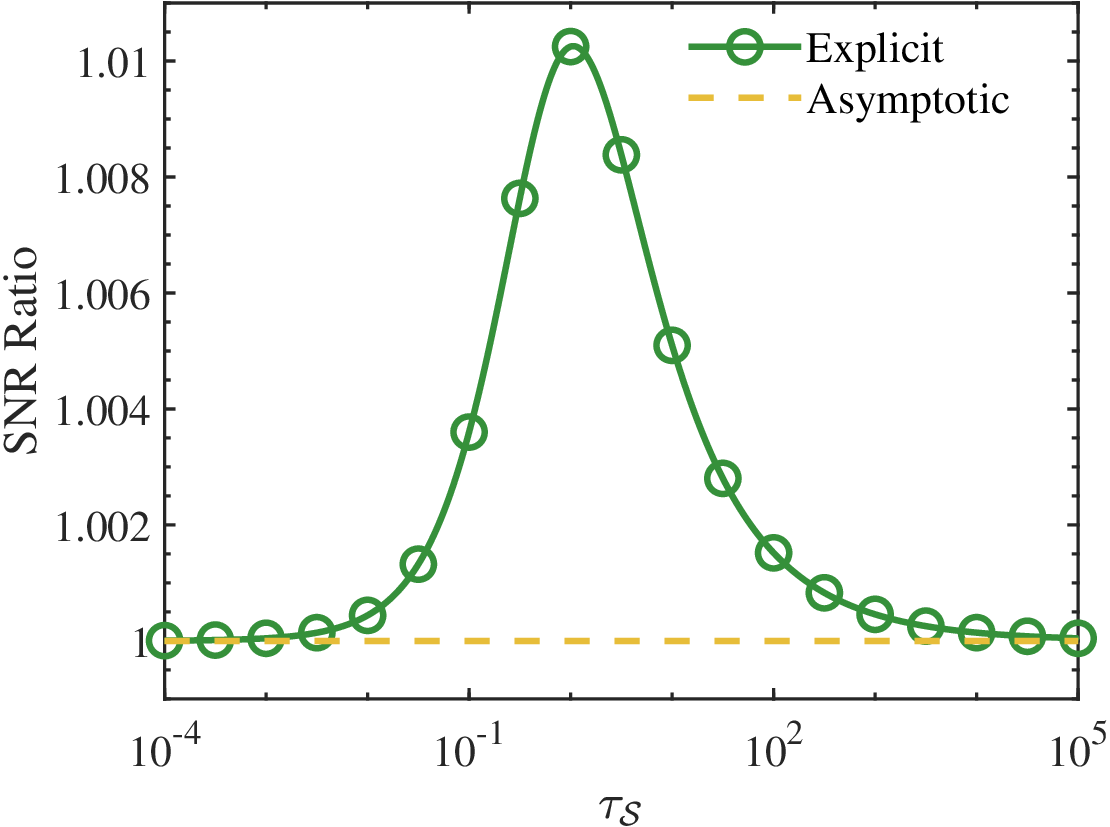}
	   \label{fig3b}	
    }
\caption{Impact of the aperture shape for LoS channels.}
    \label{Figure3}
    \vspace{-10pt}
\end{figure}

We then turn our focus to examining the impact of the aperture shape on aperture selection in LoS channels. {\figurename} {\ref{fig3a}} presents the SNRs achieved by selecting square and circular areas in terms of $\tau_{\mathcal{S}}= \frac{\lvert{\mathcal{S}}_{\mathsf{R}}\rvert}{4r^2\Psi^2}$, where the explicit SNRs are calculated using \eqref{SNR_Optimal_Aperture_Selection_Trans1} and \eqref{SNR_Optimal_Aperture_Selection_Circle}, respectively. Additionally, asymptotic results for $\tau_{\mathcal{S}}\rightarrow\infty$ and $\tau_{\mathcal{S}}\rightarrow0$ are presented, which are computed by \eqref{SNR_Optimal_Aperture_Selection_Circle_Low} (\eqref{SNR_Optimal_Aperture_Selection_Trans2_Low}) and \eqref{SNR_Optimal_Aperture_Selection_Circle_Trans1} (\eqref{SNR_Optimal_Aperture_Selection_Trans2}), respectively. These asymptotic results closely follow the explicit results as $\tau_{\mathcal{S}}$ approaches $\infty$ or $0$, which verifies our previous derivations. Furthermore, from {\figurename} {\ref{fig3a}}, we observe that selecting a circular area yields a slightly higher SNR than selecting a square one. To illustrate the performance gap further, {\figurename} {\ref{fig3b}} plots the SNR ratio between activating a circle and activating a square in terms of $\tau_{\mathcal{S}}$. It is observed that the performance gap between these two schemes is relatively small, which is consistent with our discussion in Remark \ref{Remark_Impact_Shape}.

\begin{figure}[!t]
    \centering
    \subfigbottomskip=0pt
	\subfigcapskip=-5pt
\setlength{\abovecaptionskip}{0pt}
    \subfigure[$A_x=L_x/2$.]
    {
        \includegraphics[height=0.17\textwidth]{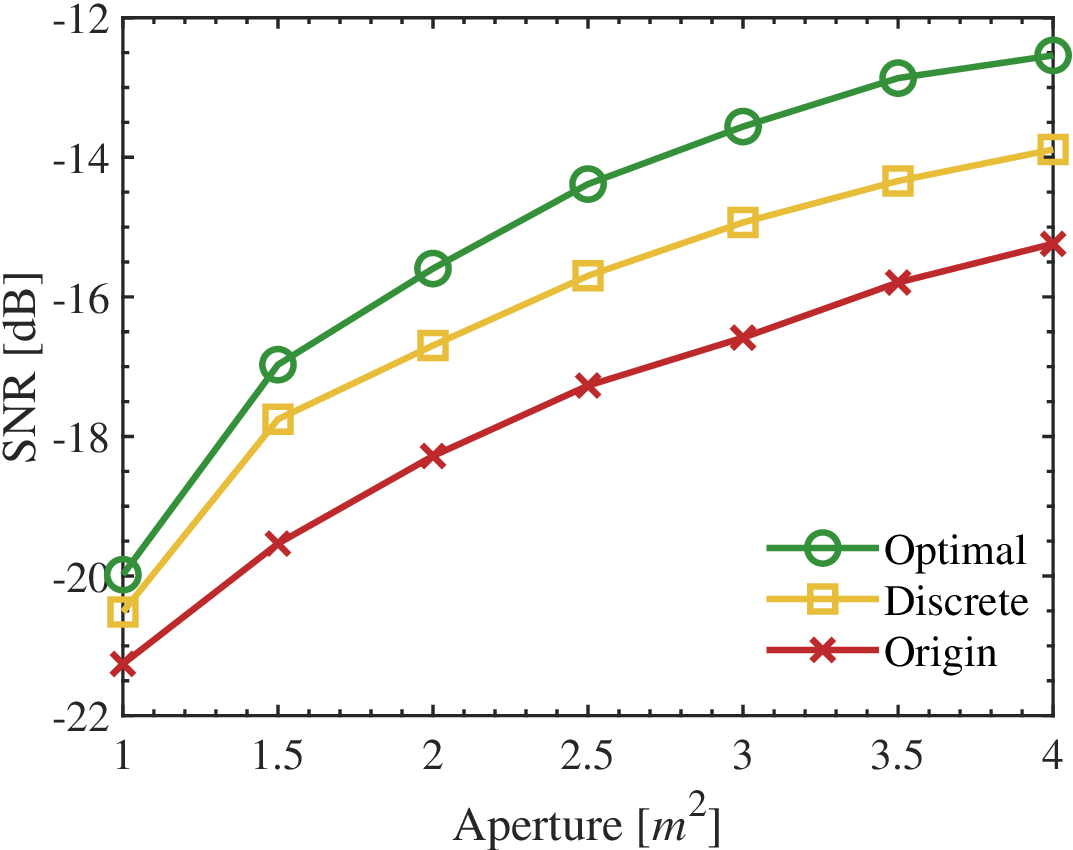}
	   \label{fig4a}	
    }
   \subfigure[OP. $\gamma_{\mathsf{th}}=-30$ dB.]
    {
        \includegraphics[height=0.17\textwidth]{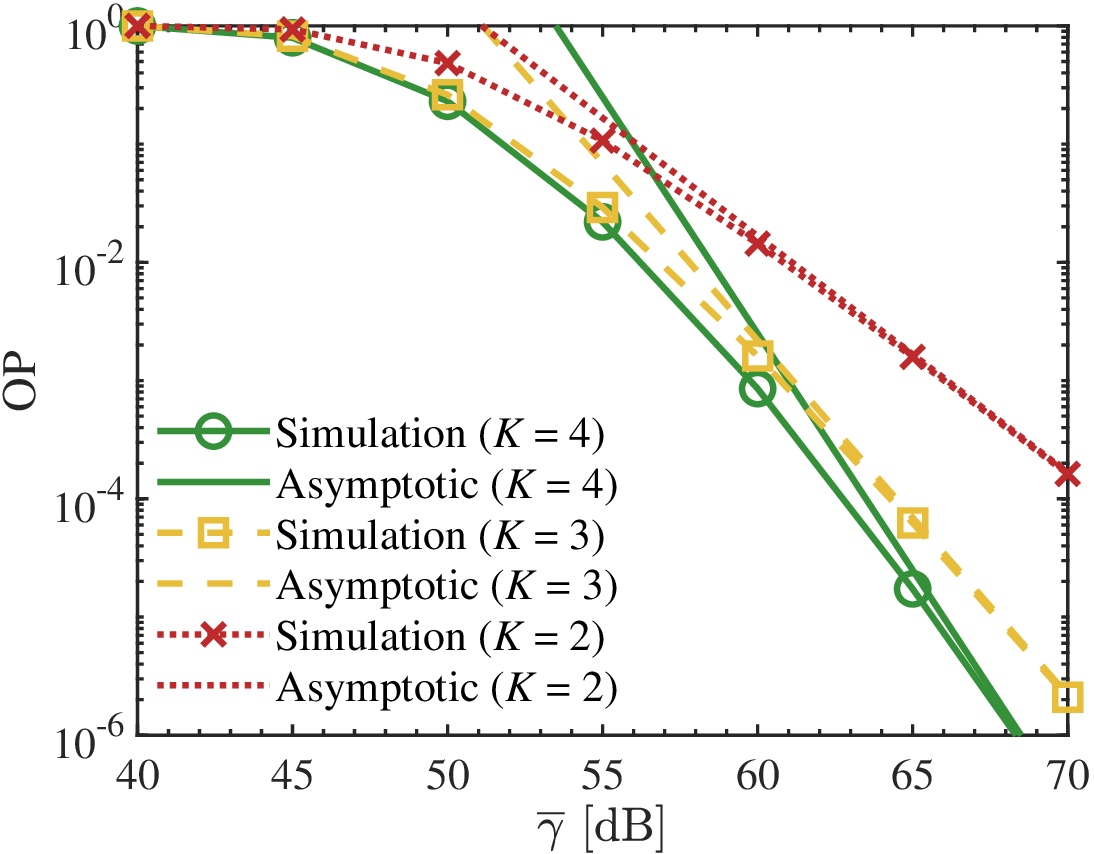}
	   \label{fig4b}	
    }
\caption{Aperture selection for NLoS channels.}
    \label{Figure4}
    \vspace{-10pt}
\end{figure}

Finally, we utilize {\figurename} {\ref{Figure4}} to discuss the performance of aperture selection under NLoS channels. Specifically, we assume there are $N_{\mathsf{s}}=4$ radonly distributed scatterers. Additionally, we set $\sigma_n^2=\frac{4\pi}{k_0^2\eta^2}$ for $n=1,\ldots,N_{\mathsf{s}}$. {\figurename} {\ref{fig4a}} illustrates the receive SNR versus $\lvert {\mathcal{A}}_{\mathsf{R}}\rvert$ when the selected aperture is an $A_x\times A_x$ square. For comparison, we present the results achieved by the optimal selection strategy (brute-force search), discrete segmentation, and no selection (${\mathcal{S}}_{\mathsf{R}}$ centered at the origin). Concerning the discrete aperture segmentation, we consider $K=5$ segments centered at ${\mathbbmss{o}}$. As observed, the proposed aperture selection methods effectively enhance the receive SNR. {\figurename} {\ref{fig4b}} further demonstrates the OP in terms of $\overline{\gamma}$ for various values of $K$ ($K\leq N_{\mathsf{s}}$), each corresponding to a random subset of ${\mathbbmss{o}}$. In the high-SNR regime, the curves for explicit OP are parallel to the asymptotic ones ($\overline{\gamma}^{-{\mathsf{r}}}\gamma_{\mathsf{th}}^{\mathsf{r}}/{{{\det}^{*}}({\mathbf{R}})}$). This indicates that the achievable diversity order obtained in Theorem \ref{Theorem_Correlation_Outage_Probability} is accurate. Since we examine the case where $K\leq N_{\mathsf{s}}$, the diversity order is constrained by $K$, which is consistent with the results shown in {\figurename} {\ref{fig4b}}.

\section{Conclusion} 
We have analyzed the receive SNR achieved by aperture selection in an uplink CAPA-based system. For LoS channels, we discussed the impact of both the size and shape of the selected aperture. For NLoS channels, we obtained the relationship between diversity order, scattering extent, and selection granularity. Our findings underscore the effectiveness of aperture selection in enhancing communication performance for both scenarios, which serves as a promising low-complexity transmission paradigm for future CAPA systems.
\begin{appendix}
\subsection{Proof of Lemma \ref{Lemma_Noise_Distribution}}\label{Proof_Lemma_Noise_Distribution}
The mean of $\int_{{\mathcal{S}}_{\mathsf{R}}}{\mathsf{h}}^{\mathsf{H}}(\mathbf{r},{\mathbf{s}}_{\mathsf{u}}){\mathsf{N}}(\mathbf{r}){\rm{d}}{\mathbf{r}}$ is calculated as
{\setlength\abovedisplayskip{2pt}
\setlength\belowdisplayskip{2pt}
\begin{equation}
\begin{split}
{\mathbbmss{E}}\left\{\int_{{\mathcal{S}}_{\mathsf{R}}}{\mathsf{h}}^{\mathsf{H}}(\mathbf{r},{\mathbf{s}}_{\mathsf{u}}){\mathsf{N}}(\mathbf{r}){\rm{d}}{\mathbf{r}}\right\}
=\int_{{\mathcal{S}}_{\mathsf{R}}}{\mathsf{h}}^{\mathsf{H}}(\mathbf{r},{\mathbf{s}}_{\mathsf{u}}){\mathbbmss{E}}\{{\mathsf{N}}(\mathbf{r})\}{\rm{d}}{\mathbf{r}},
\end{split}
\end{equation}
}which, together with the fact that ${\mathbbmss{E}}\{{\mathsf{N}}(\mathbf{r})\}=0$, yields ${\mathbbmss{E}}\{\int_{{\mathcal{S}}_{\mathsf{R}}}{\mathsf{h}}^{\mathsf{H}}(\mathbf{r},{\mathbf{s}}_{\mathsf{u}}){\mathsf{N}}(\mathbf{r}){\rm{d}}{\mathbf{r}}\}=0$. The variance is given by
{\setlength\abovedisplayskip{2pt}
\setlength\belowdisplayskip{2pt}
\begin{equation}\label{Noise_Field_Covariance}
\begin{split}
&{\mathbbmss{E}}\left\{\int_{{\mathcal{S}}_{\mathsf{R}}}{\mathsf{h}}^{\mathsf{H}}(\mathbf{r},{\mathbf{s}}_{\mathsf{u}}){\mathsf{N}}(\mathbf{r}){\rm{d}}{\mathbf{r}}
\int_{{\mathcal{S}}_{\mathsf{R}}}{\mathsf{h}}^{\mathsf{H}}(\mathbf{r}',{\mathbf{s}}_{\mathsf{u}}){\mathsf{N}}(\mathbf{r}'){\rm{d}}{\mathbf{r}}'\right\}\\
&=\int_{{\mathcal{S}}_{\mathsf{R}}}\int_{{\mathcal{S}}_{\mathsf{R}}}{\mathsf{h}}^{\mathsf{H}}(\mathbf{r},{\mathbf{s}}_{\mathsf{u}})
{\mathsf{h}}(\mathbf{r}',{\mathbf{s}}_{\mathsf{u}})
{\mathbbmss{E}}\{{\mathsf{N}}(\mathbf{r}){\mathsf{N}}(\mathbf{r}')\}{\rm{d}}{\mathbf{r}}{\rm{d}}{\mathbf{r}}'\\
&=\int_{{\mathcal{S}}_{\mathsf{R}}}{\mathsf{h}}(\mathbf{r}',{\mathbf{s}}_{\mathsf{u}})\int_{{\mathcal{S}}_{\mathsf{R}}}{\mathsf{h}}^{\mathsf{H}}(\mathbf{r},{\mathbf{s}}_{\mathsf{u}})
\sigma^2\delta(\mathbf{r}-\mathbf{r}'){\rm{d}}{\mathbf{r}}{\rm{d}}{\mathbf{r}}'.
\end{split}
\end{equation}
}Using the fact that $\int_{{\mathcal{S}}_{\mathsf{R}}}\delta({\mathbf{x}}-{\mathbf{x}}_0)f(\mathbf{x}){\rm{d}}{\mathbf{x}}=f({\mathbf{x}}_0)$, where $f(\cdot)$ is an arbitrary function defined on ${\mathcal{S}}_{\mathsf{R}}$, we obtain
{\setlength\abovedisplayskip{2pt}
\setlength\belowdisplayskip{2pt}
\begin{equation}
\eqref{Noise_Field_Covariance}=\int_{{\mathcal{S}}_{\mathsf{R}}}\sigma^2{\mathsf{h}}(\mathbf{r}',{\mathbf{s}}_{\mathsf{u}})
{\mathsf{h}}^{\mathsf{H}}(\mathbf{r}',{\mathbf{s}}_{\mathsf{u}}){\rm{d}}{\mathbf{r}}'=\sigma^2{\mathsf{a}}_{\mathsf{R}}.
\end{equation}
}Lemma \ref{Lemma_Noise_Distribution} is thus proved.
\subsection{Proof of Lemma \ref{Lemma_Received_SNR_Rectangular}}\label{Proof_Lemma_Received_SNR_Rectangular}
As per \eqref{SPD_UPA_NFC_LoS_Model_User}, the channel gain achieved by activating the aperture ${\mathcal{S}}_{\mathsf{R}}$ can be calculated as follows:
{\setlength\abovedisplayskip{2pt}
\setlength\belowdisplayskip{2pt}
\begin{equation}\label{SU_NFC_Channel_Gain}
\begin{split}
{\mathsf{a}}_{\mathsf{R}}=\int_{-\frac{A_z}{2}}^{\frac{A_z}{2}}\int_{-\frac{A_x}{2}}^{\frac{A_x}{2}}
\frac{{k_0^2\eta^2}/{(4\pi)^{2}}r{\Psi}{\rm{d}}x{\rm{d}}z}{((x-r_{x,\Phi})^2+r^2{\Psi}^2+(z-r_{z,\Theta})^2)^{\frac{3}{2}}},
\end{split}
\end{equation}
}where $r_{x,\Phi}\triangleq r\Phi-r_x$ and $r_{z,\Theta}\triangleq r\Theta-r_z$. Through employing the integral identities \cite[Eq. (2.264.5)]{gradshteyn2014table} and \cite[Eq. (2.284)]{gradshteyn2014table}, the final results follow immediately.
\subsection{Proof of Lemma \ref{Lemma_Optimal_Position_Rectangular}}\label{Proof_Lemma_Optimal_Position_Rectangular}
We use \eqref{SU_NFC_Channel_Gain} to prove this theorem, as the variables $(x,z)$ therein are decoupled. Given $z$, we have
{\setlength\abovedisplayskip{2pt}
\setlength\belowdisplayskip{2pt}
\begin{align}
\left.{\mathsf{a}}_{\mathsf{R}}\right|_z&=\int_{r_x-\frac{A_x}{2}}^{r_x+\frac{A_x}{2}}
\frac{{k_0^2\eta^2}/{(4\pi)^{2}}r{\Psi}{\rm{d}}x{\rm{d}}z}{((x-r\Phi)^2+r^2{\Psi}^2+(z-r\Theta)^2)^{\frac{3}{2}}}\\
&=\frac{k_0^2\eta^2}{(4\pi)^{2}}\frac{r{\Psi}(f_z(-r_{x,\Phi})+f_z(r_{x,\Phi}))}{r^2{\Psi}^2+(z-r\Theta)^2},
\end{align}
}where $r_{x,\Phi}\triangleq r\Phi-r_x$, $r_{z,\Theta}\triangleq r\Theta-r_z$, $f_z(x)=\frac{{A_x}/{2}+x}{(({A_x}/{2}+x)^2+r^2{\Psi}^2+(z-r\Theta)^2)^{{1}/{2}}}$. Note that $f_z(-r_{x,\Phi})+f_z(r_{x,\Phi})$ is an even function of $r_{x,\Phi}$, and thus
{\setlength\abovedisplayskip{2pt}
\setlength\belowdisplayskip{2pt}
\begin{equation}
\begin{split}
f_z(-r_{x,\Phi})+f_z(r_{x,\Phi})=f_z(-\lvert r_{x,\Phi}\rvert)+f_z(\lvert r_{x,\Phi}\rvert).
\end{split}
\end{equation}
}Furthermore, given an arbitrary real number $b$, we have
{\setlength\abovedisplayskip{2pt}
\setlength\belowdisplayskip{2pt}
\begin{equation}\label{Partial_Derviate}
\begin{split}
&\frac{{\rm{d}}}{{\rm{d}}x}\left(\frac{\frac{A_x}{2}+x}{((\frac{A_x}{2}+x)^2+b^2)^{\frac{1}{2}}}+\frac{\frac{A_x}{2}-x}{((\frac{A_x}{2}-x)^2+b^2)^{\frac{1}{2}}}\right)\\
&=\frac{-b^2}{((\frac{A_x}{2}-x)^2+b^2)^{\frac{3}{2}}}+\frac{b^2}{((\frac{A_x}{2}+x)^2+b^2)^{\frac{3}{2}}}\leq0
\end{split}
\end{equation}
}for $x>0$. Since $r^2{\Psi}^2+(z-r\Theta)^2\geq0$, we conclude from \eqref{Partial_Derviate} that $f_z(-\lvert r_{x,\Phi}\rvert)+f_z(\lvert r_{x,\Phi}\rvert)$ is monotone decreasing with $\lvert r_{x,\Phi}\rvert$. Taken together, for a given $z$, $\left.{\mathsf{a}}_{\mathsf{R}}\right|_z$ decreases with $\lvert r_{x,\Phi}\rvert$. Following the similar steps, we can also prove from \eqref{SU_NFC_Channel_Gain} that for a given $x$, the integrated function decreases with $\lvert r_{z,\Theta}\rvert$. Since the variables $(x,z)$ in \eqref{SU_NFC_Channel_Gain} are decoupled, we conclude that the definite integral in \eqref{SU_NFC_Channel_Gain} is maximized when $\lvert r_{z,\Theta}\rvert$ and $\lvert r_{x,\Phi}\rvert$ are both minimized. The final results follow immediately, and Lemma \ref{Lemma_Optimal_Position_Rectangular} is thus proved.
\end{appendix}
\bibliographystyle{IEEEtran}
\bibliography{mybib}
\end{document}